\documentclass{ifacconf}

\usepackage{graphicx}      
\usepackage{natbib}        

\usepackage{amssymb}
\usepackage{bm}
\usepackage[cmex10]{amsmath}
\usepackage{nicefrac}
\usepackage{pgfplots}
\pgfplotsset{compat=newest}
\usepackage{mathtools}
\usepackage{tikz}
\usetikzlibrary{shapes,arrows,decorations,backgrounds,positioning,calc,patterns,decorations.pathmorphing,decorations.markings,spy,backgrounds}
\usepackage{import}
\newlength\figureheight
\newlength\figurewidth

\newlength{\hatchspread}
\newlength{\hatchthickness}
\newlength{\hatchshift}
\newcommand{\hatchcolor}{}
\tikzset{hatchspread/.code={\setlength{\hatchspread}{7pt}},
	hatchthickness/.code={\setlength{\hatchthickness}{#1}},
	hatchshift/.code={\setlength{\hatchshift}{#1}},
	hatchcolor/.code={\renewcommand{\hatchcolor}{#1}}}
\tikzset{hatchspread=3pt,
	hatchthickness=0.4pt,
	hatchshift=0pt,
	hatchcolor=black}
\pgfdeclarepatternformonly[\hatchspread,\hatchthickness,\hatchshift,\hatchcolor]
{custom north east lines}
{\pgfqpoint{\dimexpr-2\hatchthickness}{\dimexpr-2\hatchthickness}}
{\pgfqpoint{\dimexpr\hatchspread+2\hatchthickness}{\dimexpr\hatchspread+2\hatchthickness}}
{\pgfqpoint{\dimexpr\hatchspread}{\dimexpr\hatchspread}}
{
	\pgfsetlinewidth{\hatchthickness}
	\pgfpathmoveto{\pgfqpoint{\dimexpr\hatchshift-0.15pt}{-0.15pt}}
	\pgfpathlineto{\pgfqpoint{\dimexpr\hatchspread+0.15pt}{\dimexpr\hatchspread-\hatchshift+0.15pt}}
	\ifdim \hatchshift > 0pt
	\pgfpathmoveto{\pgfqpoint{-0.15pt}{\dimexpr\hatchspread-\hatchshift-0.15pt}}
	\pgfpathlineto{\pgfqpoint{\dimexpr\hatchshift+0.15pt}{\dimexpr\hatchspread+0.15pt}}
	\fi
	\pgfsetstrokecolor{\hatchcolor}
	\pgfusepath{stroke}
}

\begin{document}
\begin{frontmatter}

\title{Distributed MPC: Guaranteeing Global Stability from Locally Designed Tubes\thanksref{sponsor}} 

\thanks[sponsor]{Work supported by the Harry Nicholson PhD Scholarship, Department of Automatic Control \& Systems Engineering,
University of Sheffield, and Doctoral Scholarship from CONICYT--PFCHA/ Concurso para Beca de Doctorado en el Extranjero--72150125.}

\author[Uni]{Bernardo Hernandez}
\author[Uni]{Pablo Baldivieso}
\author[Uni]{Paul Trodden} 

\address[Uni]{Department of Automatic Control \& Systems Engineering,\\ University of Sheffield, Sheffield S1 3JD, UK \\(e-mail \{prbaldivieso1, bahernandezvicente1, p.trodden\}@sheffield.ac.uk)}

\begin{abstract}                
This paper studies a fundamental relation that exists between stabilizability assumptions usually employed in distributed model predictive control implementations, and the corresponding notions of invariance implicit in such controllers. The relation is made explicit in the form of a theorem that presents sufficient conditions for global stabilizability. It is shown that constraint admissibility of local robust controllers is sufficient for the global closed-loop system to be stable, and how these controllers are related to more complex forms of control such as tube-based distributed model predictive control implementations.
\end{abstract}


\end{frontmatter}

\section{Introduction}
There exists various physical systems for which a centralized control implementation is not a suitable solution \citep{Mayne2014}. The two main reasons for this are: (i) the physical system is spread over a wide area, which makes communication to a central hub expensive and prompt to data loss (power networks, traffic networks, etc.) \citep{Baillieul2007}, (ii) the physical system is a composition of individual subsystems with clearly defined physical boundaries (autonomous vehicles, swarms of robots, etc). In the context of model-based predictive controllers, there is a third situation in which centralized control is not a viable solution, and it has to do with computational requirements. Model predictive control (MPC) is a mature control technique that provides safe and stabilizing control under appropriate design \citep{Mayne2000}. However, in order to compute the control action to be applied to the plant, it needs to solve an optimization problem at each time instant. Clearly then, plants with fast dynamics and a large number of control variables are out of the scope of standard centralized MPC implementations. A natural solution to this problem is to split the plant into smaller subsystems, and then to design local controllers. However, this creates residual interaction between different subsystems, which must be taken into account if control guarantees are to be delivered.

Many distributed MPC (DMPC) implementations have been devised to tackle this problem (for a detailed review see \cite{Scattolini2009,Christofides2013}). The main aspect in which they differ is in which type of interaction is allowed between subsystems, and how this is treated. In recent years, robust MPC techniques such as tube MPC (TMPC) \citep{Mayne2005} and others (for example \cite{Chisci2001}), have inspired several DMPC approaches \citep{Farina2012a,Riverso2012,Farina2014,Lucia2015,Trodden2016,Baldivieso2016,Hernandez2016}. The main idea of such approaches is to treat the interactions between subsystems as local disturbances that may be handled (conservatively) by robust local controllers. Communication and iteration between subsystems is then used in different ways to reduce the conservativeness induced by use of a robust method. All of these approaches, either implicitly or explicitly, rely on stabilizability assumptions. These usually require the existence of a global linear feedback with a certain structure affine to the interaction pattern of the global system. This type of assumption is also present in DMPC techniques that are not based on robust approaches such as \cite{Conte2016,Maestre2011}.

In this paper, a fundamental relation that exists between this type of stabilizability assumption, and the various concepts of invariance employed by (tube-based) DMPC architectures, is made explicit. To the best of the authors' knowledge, this relation has not been studied explicitly before (although some related results have been published, which we survey in Section 3), albeit the global stabilizability assumption is present (either explicitly or implicitly) in many of the DMPC approaches proposed to date. First, a collection of linear local robust controllers is introduced in order to study the stabilizability of the global plant from a decentralized perspective. The main result states that a sufficient condition for global stabilizability is that these local controllers are locally stabilizing and constraint admissible. Although very simple, the proposed controllers sit at the core of many tube-based DMPC architectures, which makes this result relevant to the state of the art in DMPC.

The remainder of the paper is organized as follows: Section \ref{sec.2} defines the preliminaries and standard assumptions found in many tube-based DMPC implementations. In order to contextualize the results given in this paper, Section \ref{sec.3} discusses two approaches that are commonly used to meet the previously stated assumptions. The main result is given in Section \ref{sec.4}, and three examples are presented in Section \ref{sec.5} in order to illustrate the result, and highlight its advantages and limitations. Finally, Section \ref{sec.6} provides some conclusions.

\subsubsection{Notation.}
The set of consecutive integers $\left\{1,\ldots,M\right\}$ is denoted by $\mathcal{M}$. A block-diagonal matrix $S$ with blocks $S_{i}$ and $i\in\mathcal{M}$ is denoted as $\text{diag}_{\mathcal{M}}\left(S_{i}\right)$. A vector $x$ that is formed by vertically stacked vectors $x^{i}$ with $i\in\mathcal{M}$ is denoted as $\bm{x}=\text{col}_{\mathcal{M}}\left(x^{i}\right)$. If $S$ is a matrix, $\rho\left(S\right)$ is the spectral radius of $S$; if $\rho\left(S\right)< 1$ then S is Schur. If $\mathcal{A}$ and $\mathcal{B}$ are sets, then $\mathcal{A}\oplus \mathcal{B}\coloneqq\left\{c=a+b\:|\:\forall a\in \mathcal{A},b\in \mathcal{B}\right\}$ and $\mathcal{A}\ominus \mathcal{B}\coloneqq\left\{a\in \mathcal{A}\:|\:\forall b\in \mathcal{B}, a+b\in \mathcal{A}\right\}$ (Minkowski sum and Pontryagin difference respectively). If $\mathbb{S}_{i}$ is a collection of sets with $i\in\mathcal{M}$, then $\bigoplus_{i}\mathbb{S}_{i}=\mathbb{S}_{1}\oplus\cdots\oplus\mathbb{S}_{M}$. The matrix $I_{n}$ is the identity of dimension $n$. If $x_{t}$ is the value of $x$ at time $t$, then $x_{k/t}$ is the prediction of $x_{k}$ made at time $t\leq k$.
 
\section{Preliminaries} \label{sec.2}
\subsection{Local and global dynamics}
First consider the general problem of regulating a network of $M$, possibly heterogeneous, linear time invariant (LTI) systems that interact with each other via states and inputs. For all $i\in\mathcal{M}$ the dynamics of subsystem $i$ are represented by the following state space model
\begin{equation*} 
x^{i}_{t+1}=A_{ii}x^{i}_{t}+B_{ii}u^{i}_{t}+\sum_{j\in\mathcal{N}_{i}}\left(A_{ij}x^{j}_{t}+B_{ij}u^{j}_{t}\right),
\end{equation*}
where $x^{i(j)}_{t}\in\mathbb{R}^{n_{i(j)}}$ and $u^{i(j)}_{t}\in\mathbb{R}^{m_{i(j)}}$ are the state and input vectors belonging to subsystem $i(j)$ at time $t$, with matrices $\left(A_{ij},B_{ij}\right)$ of corresponding dimension. The set $\mathcal{N}_{i}\subset\mathcal{M}$ is referred to as the set of neighbours of subsystem $i$, and contains the indexes of all the other subsystems that affect the dynamics of $i$. Given the assumed non-centralized nature of the plant, consider the case in which each subsystem $i$ is subject to local constraints of the form
\begin{equation} \label{eq.2}
x^{i}\in\mathbb{X}_{i}\subset\mathbb{R}^{n_{i}},\quad u^{i}\in\mathbb{U}_{i}\subset\mathbb{R}^{m_{i}}.
\end{equation}
\begin{assum}[Local behaviour] \label{ass.1}
For all $i\in\mathcal{M}$, the pair $\left(A_{ii},B_{ii}\right)$ is stabilizable and the set $\left(\mathbb{X}_{i}\times\mathbb{U}_{i}\right)$ is compact, convex and contains the origin in its interior.
\end{assum}

To simplify the analysis, assume that the subsystems do not share states and/or inputs. The collection of local models forms the following global constrained LTI model:
\begin{equation*} 
\bm{x}_{t+1}=\bm{A}\bm{x}_{t}+\bm{B}\bm{u}_{t},
\end{equation*}
where $\left(\bm{x},\bm{u}\right)=\left(\text{col}_{\mathcal{M}}\left(x^{i}\right),\text{col}_{\mathcal{M}}\left(u^{i}\right)\right)\in\mathbb{X}\times\mathbb{U}$, matrices $\bm{A}$ and $\bm{B}$ are compose by the blocks $\left(A_{ij},B_{ij}\right)$, and
\begin{alignat*}{2}
\mathbb{X}&=\mathbb{X}_{1}\times\cdots\times\mathbb{X}_{M}\subseteq\mathbb{R}^{n}\\
\mathbb{U}&=\mathbb{U}_{1}\times\cdots\times\mathbb{U}_{M}\subseteq\mathbb{R}^{m}.
\end{alignat*}

\subsection{Distributed control}
One of the main questions in distributed control is how to guarantee certain properties at the global scale (stability, constraint satisfaction, etc.), from a possibly distributed design. In order to achieve these global properties, many DMPC implementations employ synthesis procedures that are either centralized \citep{Maestre2011,Lucia2015}, or require the solution of problems that can be computationally expensive \citep{Conte2016,Kern2013}. In this context, and in particular with tube-based DMPC controllers, the following assumption is usually required (see for example \cite{Farina2012a,Riverso2012,Baldivieso2016,Hernandez2016}),
\begin{assum}[Block-diagonal stabilizability] \label{ass.2}
There exists a collection of local linear feedbacks $K_{i}$ such that $F_{ii}=A_{ii}+B_{ii}K_{i}$ is Schur for all $i\in\mathcal{M}$ and $\bm{F}=\bm{A}+\bm{B}\bm{K}$ is Schur, with $\bm{K}=\text{diag}_{\mathcal{M}}\left(K_{i}\right)$.
\end{assum}

Assumption \ref{ass.2} demands the existence of a globally stabilizing linear feedback $\bm{K}$ with a block-diagonal structure. This definitively limits the class of systems that can be controlled with such techniques, but a more pressing issue in the context of decentralized control, is that searching for a collection of local feedbacks that fulfils Assumption \ref{ass.2} usually requires centralized computations. There is a vast literature dedicated to analysing the impact that \emph{naive} local control design has over global behaviour (see for example \cite{Cui2002}), but perhaps an example is enough to clarify the problem.
\begin{exmp} \label{exmp.1}
Consider the coupled integrators
\begin{equation*}
\begin{bmatrix} x^1_{t+1} \\ x^2_{t+1} \end{bmatrix} = \begin{bmatrix} 1 & 0 \\ 0 & 1 \end{bmatrix}\begin{bmatrix} x^1_{t} \\ x^2_{t} \end{bmatrix} + \begin{bmatrix} 1 & 1/2 \\ 1/2 & 1 \end{bmatrix}\begin{bmatrix} u^1_{t} \\ u^2_{t} \end{bmatrix}.
\end{equation*}
The local linear feedbacks $K_{1}=K_{2}=\nicefrac{-3}{2}$ result in $\rho\left(F_{11}\right)=\rho\left(F_{22}\right)=\nicefrac{1}{2}$, however, $\rho\left(\bm{F}\right)=\nicefrac{5}{4}$.
\end{exmp}

Example \ref{exmp.1} makes clear that careful (perhaps centralized) design is required to meet Assumption \ref{ass.2}. The main result of this paper is to make explicit the fundamental relation that exists between the stabilizability requirements in Assumption \ref{ass.2} and the standard notions of invariance that are usually implicit in DMPC controllers. To aid this, various concepts of invariance are now defined.
\begin{defn}[Positive invariant (PI) set] \label{def.1}
A set $\mathbb{X}_{p}$ is said to be a PI set for the dynamics $x_{t+1}=Fx_{t}$ if $F\mathbb{X}_{p}\subseteq \mathbb{X}_{p}$.
\end{defn}
\begin{defn}[Robust positive invariant (RPI) set] \label{def.3}
A set $\mathbb{Z}$ is said to be a RPI set for the disturbed dynamics $x_{t+1}=Fx_{t}+w_{t}$ and disturbance $w\in\mathbb{W}$ if $F\mathbb{Z}\oplus\mathbb{W}\subseteq\mathbb{Z}$.
\end{defn}
\begin{rem} \label{rem.4}
Non-compact or singleton invariant sets are not considered a valid solution in the following analysis.
\end{rem}
\begin{rem} \label{rem.3}
In view of Remark \ref{rem.4}, unstable closed-loop dynamics do not accept invariant sets as described by Definitions \ref{def.1} and \ref{def.3}.
\end{rem}

\section{Global stabilization: existing approaches}  \label{sec.3}
The task of finding local linear feedbacks that fulfil Assumption \ref{ass.2} (or a similar one) appears in different steps of the implementation of DMPC controllers. Some of the techniques used to compute these feedbacks also exploit notions of invariance, and are implicitly related to the more fundamental result shown in this paper. In order to put this result into context, two of these approaches are briefly discussed.

\subsection{Linear matrix inequalities}
In the general DMPC framework (not necessarily tube-based approaches), a common practice is to tackle the distributed controller synthesis problem from a centralized perspective, and to pose a set of LMIs whose solution(s) provides an adequate candidate for the required controller (these LMIs represent standard Lyapunov stability conditions in the context of MPC). In \cite{Maestre2011} a global linear feedback, which is forced to be zero wherever there is no dynamical coupling, is proposed to play the role of the local terminal controller often used in the MPC framework \citep{Mayne2005}. In order to find this feedback, a set of local LMIs is posed alongside with a single global LMI of dimension $4n\times4n$ (where n is the overall dimension of the plant). A similar problem is found in \cite{Conte2016}, where a distributed optimization approach is proposed to find the solution to the local LMIs, in the presence of a system-wide coupled LMI.

The problem of finding \emph{separable} PI sets is tackled in a similar way in \cite{Nilsson2015}, where a set of LMIs is proposed to find simultaneously a set of local independent feedbacks $K_{i}$ that fulfil Assumption \ref{ass.2}, and a corresponding collection of joint PI sets. This procedure is proposed to tackle disturbed local dynamics, but even when no disturbance is considered, the smaller LMI is of dimension $2n\times2n$.

\subsection{Positively invariant families of sets}
Although not explicitly stated, the concept of positively invariant families of sets (PIFs), introduced by \cite{Rakovic2010}, can also be used to analyse the link between local and global closed-loop behaviour (in a centralized fashion). The concept of PIFs is more general than Definition \ref{def.1} but it is based on the same invariance properties. The main advantage of the specific parametrization proposed in \cite{Rakovic2010} is that, in the context of global to local dynamics, the dimension of the problem can be considerably reduced (from $n$ to $M$) by looking at what the authors call a \emph{comparison} system.

It is shown in \cite{Rakovic2014} that: (i) a stable comparison system is a sufficient condition for the true system to accept a PIFs, and (ii) if a system admits a PIFs, then it admits a PI set, which therefore means that the closed-loop dynamics are stable. In this context, it might prove \emph{easier} to find a stable comparison system (of dimension $M$) and then relate it to the global system in closed loop form with a particularly structured $\bm{K}$.

This notion is the underlying idea of the approach proposed in \cite{Kern2013}, where local LMIs are constructed in order to find, in a non-centralized fashion, a collection of local gains $K_{i}$ that fulfils Assumption \ref{ass.2}, and a corresponding PIFs.

In all of the these approaches, there is no guarantee that a solution does exist for the proposed set of LMIs (or stable comparison system). This should not be a surprise, given that the successful synthesis of a non-centralized controller depends greatly on the size of the interaction between neighbouring subsystems, and how these are dealt with (communication, iterative optimization, etc.).

\section{Global stabilization via tubes}  \label{sec.4}
The fundamental result shown in this paper stems from the invariance notions implicit in the robust control technique known as Tube MPC \citep{Mayne2005}, and its application to distributed control.

\subsection{Decentralized linear robust control (inside the tube)} \label{sec.1}
Suppose that subsystem $i$ has no means of obtaining information about what its neighbours' plans are. A sensible, yet conservative, way of performing non-centralized control is to view the dynamical interaction between subsystems as disturbances that must be rejected. In order to move forward, the following assumption is required.
\begin{assum} \label{ass.4}
State and input constraints are satisfied by all subsystems: $x^{i}_{t}\in\mathbb{X}_{i}$ and $u^{i}_{t}\in\mathbb{U}_{i}$ for all $i\in\mathcal{M},t\geq 0$.
\end{assum}
If Assumption \ref{ass.1} is fulfilled, then the following \emph{disturbance} sets are compact for all $i\in\mathcal{M}$
\begin{equation} \label{eq.10}
\mathbb{W}_{i}=\bigoplus_{j\in\mathcal{N}_{i}}\left(A_{ij}\mathbb{X}_{j}\oplus B_{ij}\mathbb{U}_{j}\right)=\bigoplus_{j\in\mathcal{N}_{i}}\mathbb{W}_{ij}.
\end{equation}
Moreover, if Assumption \ref{ass.4} is fulfilled, the sets $\mathbb{W}_{i}$ represent a suitable bound for the interaction between subsystems:
\begin{equation*} 
w^{i}_{t}=\sum_{j\in\mathcal{N}_{i}}\left(A_{ij}x^{j}_{t}+B_{ij}u^{j}_{t}\right)\in\mathbb{W}_{i},\quad \forall i\in\mathcal{M},t\geq 0.
\end{equation*}
In view of this and Definition \ref{def.3}, for any linear feedback $K_{i}$ that renders $F_{ii}$ Schur, there exists a compact RPI set $\mathbb{Z}_{i}$ for the local disturbed dynamics
\begin{equation} \label{eq.5.1}
x^{i}_{t+1}=F_{ii}x^{i}_{t}+w^{i}_{t},\quad w^{i}_{t}\in \mathbb{W}_{i}.
\end{equation}
Given $x^{i}_{0}\in\mathbb{Z}_{i}$ and the invariance of $\mathbb{Z}_{i}$, it follows that
\begin{equation*} 
x^{i}_{t}\in\mathbb{Z}_{i},\quad u^{i}_{t}\in K_{i}\mathbb{Z}_{i},\quad\forall t\geq 0, w^{i}_{t}\in \mathbb{W}_{i} .
\end{equation*}
\begin{prop} \label{prop.2}
If for all $i\in\mathcal{M}$ (i) $\mathbb{Z}_{i}\subset\mathbb{X}_{i}$ and $K_{i}\mathbb{Z}_{i}\subset\mathbb{U}_{i}$, and (ii) $x^{i}_{0}\in\mathbb{Z}_{i}$, then $x^{i}_{t}\in\mathbb{X}_{i}$ and $u^{i}_{t}\in\mathbb{U}_{i}$ for all $i\in\mathcal{M}$ and for all $t\geq 0$.
\end{prop}
\begin{rem}
In the context of TMPC, the sets $\mathbb{Z}_{i}$ are the cross section of the local tubes.
\end{rem}

Proposition \ref{prop.2} provides a decentralized perspective to the problem of finding stabilizing and constraint admissible local linear feedbacks. Indeed, as long as the hypotheses of Proposition \ref{prop.2} are met (for all $i\in\mathcal{M}$), independent RPI sets could be designed for each subsystem. The closed-loop global system would then take the form $\bm{F}=\bm{A}+\bm{B}\bm{K}$, which as shown in Example \ref{exmp.1}, could be unstable. This seems to defy the invariance properties of the collection of sets $\mathbb{Z}_{i}$.

In spite of the local invariance arguments, the analysis of the local dynamics does evidence an unexpected behaviour. Consider a network with $M=2$ and suppose that $K_{i}$ are indeed chosen such that $F_{ii}$ are Schur and that the hypotheses of Proposition \ref{prop.2} are met, but $\bm{F}$ is not Schur. Given $\bm{x}_{0}\in\mathbb{Z}=\mathbb{Z}_{1}\times\mathbb{Z}_{2}\subset\mathbb{X}$ and unstable closed-loop global dynamics, there must exist a $t_{1}>0$ such that, $z^{1}_{t_{1}-1}\in\mathbb{Z}_{1}$ but $z^{1}_{t_{1}}\notin\mathbb{Z}_{1}$. This means that the robust invariant property of $\mathbb{Z}_{1}$ has been broken, which can only mean that the disturbances affecting subsystem 1 (see \eqref{eq.10}) have been \emph{larger} than initially assumed. A similar argument can be made for $i=2$, which prompts the following conclusion:
\begin{prop} \label{prop.1}
Suppose $K_{i}$ are chosen such that $F_{ii}$ are Schur but $\bm{F}$ is not, and that the hypotheses of Proposition \ref{prop.2} are met. Then there exists a finite time $\hat{t}>0$ such that $\bm{x}_{\hat{t}-1}\in\mathbb{Z}$ but $\bm{x}_{\hat{t}}\notin\mathbb{X}$ or $\bm{u}_{\hat{t}}\notin\mathbb{U}$.
\end{prop}

\begin{rem} \label{rem.2}
Proposition \ref{prop.1} implies either a state \emph{jump} from inside $\mathbb{Z}$ to outside $\mathbb{X}$, or an input \emph{jump} from inside $\bm{K}\mathbb{Z}$ to outside $\mathbb{U}$. This, in some cases, might mean a discontinuity of the state trajectories, which is not characteristic of the type of (linear) systems being analysed.
\end{rem}

\subsection{From local robustness to global stability (main result)}
The uncharacteristic behaviour made explicit by Remark \ref{rem.2} can be explained by the fundamental relation that exists between Assumption \ref{ass.2} and the notions of invariance employed in Proposition \ref{prop.2} (and by TMPC). Consider the following definition:
\begin{defn}[Admissibility of tube $i$] \label{def.2}
A tube $i$ corresponding to a particular linear feedback $K_{i}$ is said to be constraint admissible if the first part of the hypothesis of Proposition \ref{prop.2} holds, i.e., if $\mathbb{Z}_{i}\subset\mathbb{X}_{i}$ and $K_{i}\mathbb{Z}_{i}\subset\mathbb{U}_{i}$.
\end{defn}
The main result of this paper is now stated:
\begin{thm} \label{thm.2}
If there exists constraint admissible tubes for all subsystems $i\in\mathcal{M}$, then the collection of local gains $K_{i}$ related to these admissible tubes fulfils Assumption \ref{ass.2}.
\end{thm}
\begin{pf}
First, in view of Remark \ref{rem.3}, if $F_{ii}$ admits a RPI set $\mathbb{Z}_{i}$, then $F_{ii}$ is Schur. Given Definition \ref{def.3}, and standard Minkowski sum properties, it is clear that if $\mathbb{Z}_{i}$ is RPI for the disturbed dynamics in \eqref{eq.5.1}, then
\begin{equation} \label{eq.12}
F_{ii}\mathbb{Z}_{i}\oplus\mathbb{V}_{i}\subseteq\mathbb{Z}_{i},\quad\forall\mathbb{V}_{i}\subseteq\mathbb{W}_{i},\forall i\in\mathcal{M}.
\end{equation}
Moreover, for all $i,j\in\mathcal{M}$ with $i\neq j$,
\begin{subequations} \label{eq.14}
\begin{alignat}{1}
\mathbb{V}_{ij}=\left(A_{ij}+B_{ij}K_{j}\right)\mathbb{Z}_{j}&\subseteq A_{ij}\mathbb{Z}_{j}\oplus B_{ij}K_{j}\mathbb{Z}_{j}\\
&\subset A_{ij}\mathbb{X}_{j}\oplus B_{ij}\mathbb{U}_{j}=\mathbb{W}_{ij}\\
\implies\mathbb{V}_{ij}&\subset\mathbb{W}_{ij}\\
\implies\mathbb{V}_{i}=\bigoplus_{j\in\mathcal{N}_{i}}\mathbb{V}_{ij}&\subset\mathbb{W}_{i},
\end{alignat}
\end{subequations}
where the first inclusion follows from Minkowski sum properties and the second inclusion from the constraint admissibility of tube $i$. Equations \eqref{eq.12} and \eqref{eq.14} imply
\begin{equation*} 
F_{ii}\mathbb{Z}_{i}\oplus\bigoplus_{j\in\mathcal{N}_{i}}F_{ij}\mathbb{Z}_{j}\subset\mathbb{Z}_{i},\quad\forall i\in\mathcal{M},
\end{equation*}
and hence, $\mathbb{Z}=\mathbb{Z}_{1}\times\cdots\times\mathbb{Z}_{M}\subseteq\mathbb{R}^{n}$ is a PI set for the closed-loop dynamics $\bm{F}$. In view of this and Remark \ref{rem.3}, $\bm{F}$ is Schur.
\end{pf}

\begin{rem}
Theorem \ref{thm.2} presents sufficient conditions for a collection locally stabilizing linear feedbacks to be globally stabilizing.
\end{rem}
\begin{rem}
Unstable global dynamics, as demanded by Proposition \ref{prop.1}, imply Assumption \ref{ass.4} is not met, and therefore (some of) the disturbance sets are unbounded. The corresponding RPI sets are also unbounded, thus, the set inclusions required for constraint admissibility of the tubes cannot be met. This implies that the behaviour described by Remark \ref{rem.2} is not possible, because the tubes would not have been admissible.
\end{rem}
A collection of constraint admissible tubes $\mathbb{Z}_{i}$ is a family of jointly RPI sets. This collection forms what can be referred to as a \emph{square} PI set, given that $\mathbb{Z}$ is defined by the cross product of subsets of the local dynamics state space. This is also pointed out in \cite{Maestre2011,Lucia2015,Nilsson2015,Conte2016}.
\begin{rem}\label{rem.1}
The existence of such a particular PI set is a necessary condition for the tubes to be system-wide admissible. This clarifies the main source of conservativeness of Theorem \ref{thm.2}.
\end{rem}
Large constraint sets for a single subsystem might render impossible to find admissible tubes for the affected neighbours. However, Theorem \ref{thm.2} remains useful for the task of analysing local-to-global stabilizability (the constraint sets could be shrunk, similar to the approach taken in \cite{Maestre2011}).

In the parametrization of PIFs proposed by \cite{Rakovic2010}, each element in the family of sets is also \emph{square}. A marginally stable comparison system that admits the corresponding eigenvector as an initial scaling factor, also provides a \emph{square} PI, and therefore, a collection of local linear feedbacks $K_{i}$ that fulfils Assumption \ref{ass.2}.

\subsection{Tube MPC}
The main purpose of the linear robust controller presented in Section \ref{sec.1} was to show the fundamental result given by Theorem \ref{thm.2}. On its own, this controller can only guarantee robust stabilizability of its region of attraction (the set $\mathbb{Z}$). However, such a controller is at the core of many tube-based DMPC architectures. In order to make this clear, TMPC is now briefly described.

TMPC aims to solves the regulation problem for a nominal undisturbed version of the plant, while securing that the state of the true plant remains bounded inside a \emph{tube} centred (at each time) at the nominal undisturbed trajectory. Define the nominal model for subsystem $i$ as
\begin{subequations}
\begin{alignat}{1}
&\hat{x}^{i}_{t+1}=A_{ii}\hat{x}^{i}_{t}+B_{ii}\hat{u}^{i}_{t} \label{eq.6.1}\\
&\hat{x}^{i}\in\hat{\mathbb{X}}_{i},\quad \hat{u}^{i}\in\hat{\mathbb{U}}_{i},
\end{alignat}
\end{subequations}
then, the control action to be applied to the true plant, at each time step, is computed via the following control policy
\begin{equation} \label{eq.7}
u^{i}_{t}=\hat{u}^{i}_{t/t}+K_{i}\left(x^{i}_{t}-\hat{x}^{i}_{t/t}\right).
\end{equation}
The first term in \eqref{eq.7} is a stabilizing control action for the nominal model, and the second one is designed to reject the disturbances. The local feedback $K_{i}$ is any stabilizing gain for the pair $\left(A_{ii},B_{ii}\right)$. The dynamics of the trajectory deviation $z^{i}_{t}=x^{i}_{t}-\hat{x}^{i}_{t}$ are therefore defined by
\begin{equation} \label{eq.11}
z^{i}_{t+1}=F_{ii}z^{i}_{t}+w^{i}_{t},
\end{equation}
where the disturbance $w^{i}$ represents the unknown dynamical coupling, and therefore belongs to $\mathbb{W}_{i}$ as defined in \eqref{eq.10} (exact same dynamics as \eqref{eq.5.1}).

The pair $\left(\hat{u}^{i}_{t/t},\hat{x}^{i}_{t/t}\right)$ is obtained from the following open loop optimal control problem, which is standard in nominal MPC implementations,
\begin{equation*} 
\mathbb{P}_N(x^{i}_{t}):\quad\min_{\hat{x}^{i}_{t/t},\hat{u}^{i}_{k/t}}\sum_{k=t}^{t+N-1}l_{i}\left(x,u\right)+V^{i}_{f}\left(\hat{x}^{i}_{N/t}\right)
\end{equation*}
subject to the dynamics in \eqref{eq.6.1} and, for $k=t,\ldots,t+N-1$:
\begin{subequations} \label{eq.9}
	\begin{alignat}{1}
	x^{i}_{t}-\hat{x}^{i}_{t/t}&\in\mathbb{Z}_{i} \label{eq.9.1}\\
	\hat{x}^{i}_{k/t}&\in\hat{\mathbb{X}}_{i}\subseteq\mathbb{X}_{i}\ominus\mathbb{Z}_{i}\\
	\hat{u}^{i}_{k/t}&\in\hat{\mathbb{U}}_{i}\subseteq\mathbb{U}_{i}\ominus K_{i}\mathbb{Z}_{i}\\
	\hat{x}^{i}_{t+N/t}&\in\hat{\mathbb{X}}^{f}_{i}\subseteq\hat{\mathbb{X}}_{i}.
	\end{alignat}
\end{subequations}
\begin{thm} \label{thm.1}
If (i) $\mathbb{Z}_{i}$ is an RPI set for the dynamics in \eqref{eq.11} and disturbance $w^{i}\in\mathbb{W}_{i}$, (ii) $l_{i}$, $V^{i}_{f}$ and $\hat{\mathbb{X}}^{f}_{i}$ are designed following standard MPC arguments to ensure asymptotic stability of the nominal undisturbed system, and (iii) the set $\hat{\mathbb{X}}_{i}\times\hat{\mathbb{U}}_{i}$ fulfils Assumption \ref{ass.1}, then (a) the set $\mathbb{Z}_{i}$ is robustly asymptotically stable for system \eqref{eq.5.1} in closed loop with \eqref{eq.7} and (b) constraints \eqref{eq.2} are met at all times.
\end{thm}
\begin{pf}
The reader is referred to \cite{Mayne2005} for the proof.
\end{pf}
Note that, for all $t>0$, constraint \eqref{eq.9.1} may be replaced by $\hat{x}^{i}_{t/t}=\hat{x}^{i}_{t/t-1}$ (i.e., independent time evolution of the nominal system, see \cite{Rawlings2014}). Given $\bm{x}_{0}\in\mathbb{Z}$, the closed-loop dynamics reduce to $\bm{F}=\bm{A}+\bm{B}\bm{K}$, which, again, could show the behaviour described by Proposition \ref{prop.2} and Remark \ref{rem.2}. However, Theorem \ref{thm.2} shows that this is not the case.
\begin{cor}
If the hypotheses of Theorem \ref{thm.1} are met for all $i\in\mathcal{M}$, then $\bm{F}$ is Schur.
\end{cor}
\begin{pf}
Hypotheses (i) and (iii) of Theorem \ref{thm.1} are equivalent to the existence of a constraint admissible tube for subsystem $i$, hence, if met for all $i\in\mathcal{M}$, the hypothesis of Theorem \ref{thm.2} is met, and the collection of local linear feedbacks $K_{i}$ fulfils Assumption \ref{ass.2}.
\end{pf}

\section{Numerical Example} \label{sec.5}
In order to illustrate the main result of this paper, consider a global plant composed by four trucks modelled by point masses, aligned in an horizontal plane (see Fig. \ref{fig.1}). Each truck is dynamically coupled to its immediate neighbours through springs $k_{ij}$ and dampers $c_{ij}$. The control objective is to steer the whole system towards an arbitrary equilibrium point using locally applied horizontal forces. This plant is used as a numerical example for various DMPC algorithms, see for example \cite{Riverso2012,Baldivieso2016,Hernandez2016,Trodden2016}.

\tikzstyle{block} = [draw, fill=none, rectangle, minimum width=1.4cm ,minimum height=1.4cm]
\tikzstyle{wheel} = [draw, fill=black, circle, minimum size=0.2cm,inner sep=0pt]
\tikzstyle{spring}=[thick,decorate,decoration={zigzag,pre length=0.3cm,post length=0.3cm,segment length=6}]
\tikzstyle{damper}=[thick,decoration={markings,  
	mark connection node=dmp,
	mark=at position 0.5 with 
	{
		\node (dmp) [thick,inner sep=0pt,transform shape,rotate=-90,minimum width=15pt,minimum height=3pt,draw=none] {};
		\draw [thick] ($(dmp.north east)+(2pt,0)$) -- (dmp.south east) -- (dmp.south west) -- ($(dmp.north west)+(2pt,0)$);
		\draw [thick] ($(dmp.north)+(0,-5pt)$) -- ($(dmp.north)+(0,5pt)$);
	}
}, decorate]
\tikzstyle{ground}=[fill,pattern=north east lines,draw=none,minimum width=11cm,minimum height=0.3cm]
\begin{figure}
\begin{center}
\begin{tikzpicture}[scale=0.8, every node/.style={scale=0.8}]
\node at (0,0) [block, name=T1, align=center] {$m_1=3$};
\node [right= 1.2cm of T1, block, name=T2, align=center] {$m_2=2$};
\node [right= 1.2cm of T2, block, name=T3, align=center] {$m_3=3$};
\node [right= 1.2cm of T3, block, name=T4, align=center] {$m_4=6$};
\node (w1) at (T1.south west)[wheel,align=center,xshift =0.2cm,yshift=-0.1cm] {};
\node (w2) at (T1.south east)[wheel,align=center,xshift=-0.2cm,yshift=-0.1cm] {};
\node (w3) at (T2.south west)[wheel,align=center,xshift =0.2cm,yshift=-0.1cm] {};
\node (w4) at (T2.south east)[wheel,align=center,xshift=-0.2cm,yshift=-0.1cm] {};
\node (w5) at (T3.south west)[wheel,align=center,xshift =0.2cm,yshift=-0.1cm] {};
\node (w6) at (T3.south east)[wheel,align=center,xshift=-0.2cm,yshift=-0.1cm] {};
\node (w7) at (T4.south west)[wheel,align=center,xshift =0.2cm,yshift=-0.1cm] {};
\node (w8) at (T4.south east)[wheel,align=center,xshift=-0.2cm,yshift=-0.1cm] {};

\draw [spring] ($(T1.east)+(0,0.3cm)$)  to node[above,yshift=0.15cm] {\small$k_{12}=7.5$} ($(T2.west)+(0,0.3cm)$);
\draw [damper] ($(T1.east)+(0,-0.3cm)$) to node[below,yshift=-0.17cm] {\small$c_{12}=4$} ($(T2.west)+(0,-0.3cm)$);
\draw [spring] ($(T2.east)+(0,0.3cm)$)  to node[above,yshift=0.15cm] {\small$k_{23}=0.75$} ($(T3.west)+(0,0.3cm)$);
\draw [damper] ($(T2.east)+(0,-0.3cm)$) to node[below,yshift=-0.17cm] {\small$c_{23}=0.25$} ($(T3.west)+(0,-0.3cm)$);
\draw [spring] ($(T3.east)+(0,0.3cm)$)  to node[above,yshift=0.15cm] {\small$k_{34}=1$} ($(T4.west)+(0,0.3cm)$);
\draw [damper] ($(T3.east)+(0,-0.3cm)$) to node[below,yshift=-0.17cm] {\small$c_{34}=0.3$} ($(T4.west)+(0,-0.3cm)$);

\node (gr) at (T1.south west)[ground, align=center, yshift=-0.35cm, xshift=5cm] {};
\draw[-] (gr.north west) to (gr.north east);

\node (p1) at (T1.north)[xshift =0.5cm,yshift=0.3cm] {};
\node (p2) at (T2.north)[xshift =0.5cm,yshift=0.3cm] {};
\node (p3) at (T3.north)[xshift =0.5cm,yshift=0.3cm] {};
\node (p4) at (T4.north)[xshift =0.5cm,yshift=0.3cm] {};
\draw[-latex] (T1.north) |- node[above,xshift=0.15cm] {\small$u^{1}$} (p1);
\draw[-latex] (T2.north) |- node[above,xshift=0.15cm] {\small$u^{2}$} (p2);
\draw[-latex] (T3.north) |- node[above,xshift=0.15cm] {\small$u^{3}$} (p3);
\draw[-latex] (T4.north) |- node[above,xshift=0.15cm] {\small$u^{4}$} (p4);
\end{tikzpicture}
\caption{Global plant: four dynamically coupled trucks.} 
\label{fig.1}
\end{center}
\end{figure}
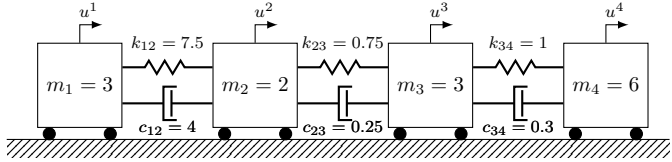

The state vector of the global plant is composed of the horizontal position and velocity of all trucks, and so $\left(\bm{x},\bm{u}\right)\in\mathbb{R}^{8}\times\mathbb{R}^{4}$. A single subsystem is associated to each truck, the local state and input vectors $\left(x^{i},u^{i}\right)$ represent the position, velocity and force of the corresponding truck. A sampling time of $T_{s}=0.1$ is used to discretize the system (the values of the matrices $\bm{A}$ and $\bm{B}$ are omitted).

\subsection{Case 1: admissible tubes}
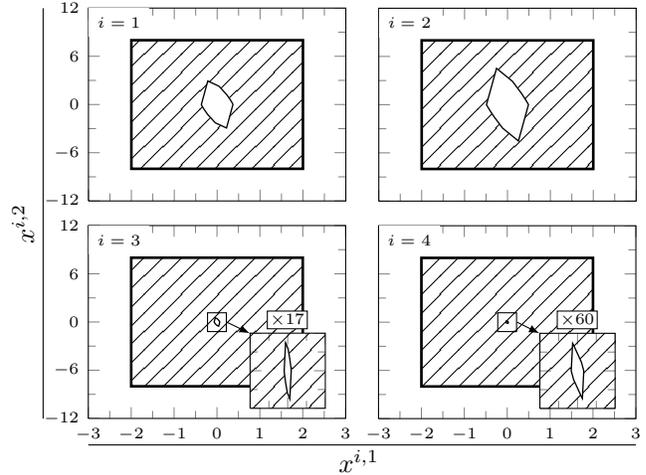
\begin{figure}
\begin{center}
\setlength\figureheight{0.3\textwidth}
\setlength\figurewidth{0.4\textwidth}
\newcommand{\picheight}{0.47}
\newcommand{\picwidth}{0.47}
\newcommand{\picwidthh}{0.53}

\begin{tikzpicture}[align=center,
square node/.style={rectangle,draw=black,fill=none,thin,minimum width=2.5mm,minimum height=2.5mm},
every pin/.style={fill=white}]
\begin{axis}[%
width=\picwidth\figurewidth,
height=\picheight\figureheight,
at={(0\figurewidth,1\figureheight)},
scale only axis,
scale only axis,
xmin=-3,
xmax=3,
xtick={-3,-2,...,3},
xticklabels={,,},
ymin=-12,
ymax=12,
ytick={-12,-6,...,12},
minor x tick num={1},
minor y tick num={1},
yticklabel style = {font=\tiny},
xticklabel style = {font=\tiny},
ylabel={$x^{i,2}$},
y label style={at={(-0.6cm,-0.3cm)}},
]
\addplot[area legend,solid,line width=1.0pt,draw=black,pattern=custom north east lines,forget plot]
table[row sep=crcr] {%
	x	y\\
	-2	-8\\
	2	-8\\
	2	8\\
	-2	8\\
}--cycle;
\label{plot:admX}
\addplot[area legend,solid,line width=0.5pt,draw=black,fill=white,forget plot,postaction={pattern=north west lines}]
table[row sep=crcr] {%
	x	y\\
   0.368448027426730  -0.000162908593819\\
   0.368448799999081  -0.000179425011670\\
   0.218828342321213  -2.921402697689580\\
   0.218663651886365  -2.924267462760100\\
  -0.041234398842243  -2.283205029442100\\
  -0.041503580976335  -2.282505004132050\\
  -0.217640499998698  -1.260309444944030\\
  -0.217821191678819  -1.259251310965580\\
  -0.308879617441970  -0.575888702299051\\
  -0.308972572682921  -0.575187820418232\\
  -0.348585120997550  -0.224382582238845\\
  -0.348625400493980  -0.224024521965303\\
  -0.363279328353610  -0.072271656277245\\
  -0.363294164042220  -0.072117369938229\\
  -0.367656350060199  -0.016330548535438\\
  -0.367660734983506  -0.016274080168849\\
  -0.368448027426730   0.000162908593819\\
  -0.368448799999081   0.000179425011669\\
  -0.218828342321213   2.921402697689580\\
  -0.218663651886365   2.924267462760100\\
   0.041234398842243   2.283205029442100\\
   0.041503580976335   2.282505004132050\\
   0.217640499998698   1.260309444944030\\
   0.217821191678819   1.259251310965580\\
   0.308879617441970   0.575888702299050\\
   0.308972572682921   0.575187820418231\\
   0.348585120997550   0.224382582238845\\
   0.348625400493980   0.224024521965303\\
   0.363279328353610   0.072271656277244\\
   0.363294164042220   0.072117369938228\\
   0.367656350060199   0.016330548535438\\
   0.367660734983506   0.016274080168849\\
}--cycle;
\label{plot:admZ}
\coordinate (pta) at (axis description cs:0,1);
\node[fill=white,align=center,anchor=north west] at (pta) {\tiny$i=1$};
\coordinate (ptl3) at (axis description cs:0,1);
\end{axis}

\begin{axis}[%
width=\picwidth\figurewidth,
height=\picheight\figureheight,
at={(\picwidthh\figurewidth,1\figureheight)},
scale only axis,
scale only axis,
xmin=-3,
xmax=3,
xtick={-3,-2,...,3},
xticklabels={,,},
ymin=-12,
ymax=12,
ytick={-12,-6,...,12},
yticklabels={,,},
minor x tick num={1},
minor y tick num={1},
yticklabel style = {font=\tiny},
xticklabel style = {font=\tiny},
]
\addplot[area legend,solid,line width=1.0pt,draw=black,pattern=custom north east lines,forget plot]
table[row sep=crcr] {%
	x	y\\
	-2	-8\\
	2	-8\\
	2	8\\
	-2	8\\
}--cycle;

\addplot[area legend,solid,line width=0.5pt,draw=black,fill=white,forget plot,postaction={pattern=north west lines}]
table[row sep=crcr] {%
	x	y\\
   0.255008543583299  -4.518314341106630\\
  -0.114932146924777  -2.911986959291830\\
  -0.328149226588979  -1.391571553857120\\
  -0.425841651857804  -0.584167470209297\\
  -0.465908525316492  -0.227063944638343\\
  -0.481241185232219  -0.083612028120101\\
  -0.486820626334139  -0.029503832205347\\
  -0.488770129674639  -0.010035232242447\\
  -0.489427456108811  -0.003297944179057\\
  -0.489641762847750  -0.001046327547677\\
  -0.489709299098649  -0.000318618898758\\
  -0.489729813750556  -0.000091391676066\\
  -0.489735783347873  -0.000023085748628\\
  -0.489737427877648  -0.000003462406248\\
  -0.255008543583298   4.518314341106630\\
   0.114932146924777   2.911986959291830\\
   0.328149226588979   1.391571553857120\\
   0.425841651857804   0.584167470209296\\
   0.465908525316492   0.227063944638342\\
   0.481241185232220   0.083612028120101\\
   0.486820626334139   0.029503832205347\\
   0.488770129674640   0.010035232242447\\
   0.489427456108811   0.003297944179057\\
   0.489641762847750   0.001046327547677\\
   0.489709299098649   0.000318618898758\\
   0.489729813750557   0.000091391676066\\
   0.489735783347874   0.000023085748627\\
   0.489737427877649   0.000003462406247\\
}--cycle;
\coordinate (ptc) at (axis description cs:0,1);
\node[fill=white,align=center,anchor=north west] at (ptc) {\tiny$i=2$};
\end{axis}

\begin{axis}[%
width=\picwidth\figurewidth,
height=\picheight\figureheight,
at={(0\figurewidth,\picheight\figureheight)},
scale only axis,
xmin=-3,
xmax=3,
xtick={-3,-2,...,3},
ymin=-12,
ymax=12,
ytick={-12,-6,...,12},
minor x tick num={1},
minor y tick num={1},
yticklabel style = {font=\tiny},
xticklabel style = {font=\tiny},
xlabel={$x^{i,1}$},
x label style={at={(3.6cm,-0.3cm)}},
]
\addplot[area legend,solid,line width=1.0pt,draw=black,pattern=custom north east lines,forget plot]
table[row sep=crcr] {%
	x	y\\
	-2	-8\\
	2	-8\\
	2	8\\
	-2	8\\
}--cycle;

\addplot[area legend,solid,line width=0.5pt,draw=black,fill=white,forget plot,postaction={pattern=north west lines}]
table[row sep=crcr] {%
	x	y\\
   0.065641451111154  -0.000523573336477\\
   0.039481013834674  -0.521884609470779\\
   0.039434215714766  -0.521770303650118\\
  -0.006981625610713  -0.407263698795151\\
  -0.007013497656042  -0.407079840812772\\
  -0.038569751770231  -0.224746470483960\\
  -0.038586278105004  -0.224623377756604\\
  -0.054934351815940  -0.102756218115168\\
  -0.054941564632338  -0.102692948204965\\
  -0.062071587462886  -0.040107545474603\\
  -0.062074268936728  -0.040080123037796\\
  -0.064722923001641  -0.012973296075579\\
  -0.064723729259905  -0.012963179544620\\
  -0.065519140783651  -0.002970870731936\\
  -0.065667513571400  -0.000007215557322\\
  -0.065641451111154   0.000523573336477\\
  -0.039481013834674   0.521884609470779\\
  -0.039434215714766   0.521770303650118\\
   0.006981625610713   0.407263698795151\\
   0.007013497656042   0.407079840812772\\
   0.038569751770231   0.224746470483960\\
   0.038586278105004   0.224623377756604\\
   0.054934351815940   0.102756218115168\\
   0.054941564632338   0.102692948204965\\
   0.062071587462886   0.040107545474603\\
   0.062074268936728   0.040080123037796\\
   0.064722923001641   0.012973296075579\\
   0.064723729259905   0.012963179544620\\
   0.065519140783651   0.002970870731936\\
   0.065667513571400   0.000007215557322\\
}--cycle;
\coordinate (ptb) at (axis description cs:0,1);
\node[fill=white,align=center,anchor=north west] at (ptb) {\tiny$i=3$};
\coordinate (ptl1) at (axis description cs:0,0);
\coordinate (pt3) at (axis cs:0,0);
\end{axis}

\begin{axis}[%
width=1cm,
height=1cm,
at={(0,0)},
scale only axis,
xmin=-0.7,
xmax=0.7,
xtick={-0.7,0,0.7},
xticklabels={,,},
ymin=-0.7,
ymax=0.7,
ytick={-0.7,0,0.7},
yticklabels={,,},
xshift=2.15cm,yshift=2.7cm,
minor x tick num={1},
minor y tick num={1},
axis background/.style={fill=white},
yticklabel style = {font=\tiny},
]
\addplot[area legend,solid,line width=1.0pt,draw=black,pattern=custom north east lines,forget plot]
table[row sep=crcr] {%
	x	y\\
	-2	-8\\
	2	-8\\
	2	8\\
	-2	8\\
}--cycle;

\addplot[area legend,solid,line width=0.5pt,draw=black,fill=white,forget plot,postaction={pattern=north west lines}]
table[row sep=crcr] {%
	x	y\\
   0.065641451111154  -0.000523573336477\\
   0.039481013834674  -0.521884609470779\\
   0.039434215714766  -0.521770303650118\\
  -0.006981625610713  -0.407263698795151\\
  -0.007013497656042  -0.407079840812772\\
  -0.038569751770231  -0.224746470483960\\
  -0.038586278105004  -0.224623377756604\\
  -0.054934351815940  -0.102756218115168\\
  -0.054941564632338  -0.102692948204965\\
  -0.062071587462886  -0.040107545474603\\
  -0.062074268936728  -0.040080123037796\\
  -0.064722923001641  -0.012973296075579\\
  -0.064723729259905  -0.012963179544620\\
  -0.065519140783651  -0.002970870731936\\
  -0.065667513571400  -0.000007215557322\\
  -0.065641451111154   0.000523573336477\\
  -0.039481013834674   0.521884609470779\\
  -0.039434215714766   0.521770303650118\\
   0.006981625610713   0.407263698795151\\
   0.007013497656042   0.407079840812772\\
   0.038569751770231   0.224746470483960\\
   0.038586278105004   0.224623377756604\\
   0.054934351815940   0.102756218115168\\
   0.054941564632338   0.102692948204965\\
   0.062071587462886   0.040107545474603\\
   0.062074268936728   0.040080123037796\\
   0.064722923001641   0.012973296075579\\
   0.064723729259905   0.012963179544620\\
   0.065519140783651   0.002970870731936\\
   0.065667513571400   0.000007215557322\\
}--cycle;
\coordinate (pt4) at (axis description cs:0,1);
\end{axis}

\node[square node, name=z2] at (pt3) {};
\draw [-latex] (z2.east)  to node[above,draw=black,fill=white,inner sep=1pt,yshift=-0.02cm,xshift=0.65cm]{\tiny$\times 17$} (pt4);

\begin{axis}[%
width=\picwidth\figurewidth,
height=\picheight\figureheight,
at={(\picwidthh\figurewidth,\picheight\figureheight)},
scale only axis,
scale only axis,
xmin=-3,
xmax=3,
xtick={-3,-2,...,3},
ymin=-12,
ymax=12,
ytick={-12,-6,...,12},
yticklabels={,,},
minor x tick num={1},
minor y tick num={1},
yticklabel style = {font=\tiny},
xticklabel style = {font=\tiny},
]
\addplot[area legend,solid,line width=1.0pt,draw=black,pattern=custom north east lines,forget plot]
table[row sep=crcr] {%
	x	y\\
	-2	-8\\
	2	-8\\
	2	8\\
	-2	8\\
}--cycle;

\addplot[area legend,solid,line width=0.5pt,draw=black,fill=white,forget plot,postaction={pattern=north west lines}]
table[row sep=crcr] {%
	x	y\\
   0.027661937069151  -0.001664166370806\\
   0.020341827123091  -0.147924141063745\\
   0.020339896351905  -0.147925685773233\\
   0.020169407013200  -0.148041392069230\\
   0.005119039886861  -0.156479944975660\\
   0.005117277613036  -0.156475036165270\\
   0.004962900261962  -0.156037595854310\\
  -0.008559296312698  -0.117079376436903\\
  -0.008560528152203  -0.117073678691818\\
  -0.008667996370438  -0.116573119254879\\
  -0.018043066683797  -0.072604509112489\\
  -0.018043792649143  -0.072600089714293\\
  -0.018106920143348  -0.072213862560532\\
  -0.023595896808368  -0.038462879577266\\
  -0.023596261370248  -0.038460070568012\\
  -0.023627847325318  -0.038215436556051\\
  -0.026364235883706  -0.016911930224764\\
  -0.026376977076731  -0.016779627729080\\
  -0.027474222591930  -0.005296642277619\\
  -0.027477358058877  -0.005236796551065\\
  -0.027742051270691  -0.000065788372291\\
  -0.027741196337047  -0.000045803952192\\
  -0.027661937069151   0.001664166370806\\
  -0.020341827123091   0.147924141063745\\
  -0.020339896351905   0.147925685773233\\
  -0.020169407013200   0.148041392069230\\
  -0.005119039886861   0.156479944975660\\
  -0.005117277613036   0.156475036165270\\
  -0.004962900261962   0.156037595854310\\
   0.008559296312698   0.117079376436903\\
   0.008560528152203   0.117073678691818\\
   0.008667996370438   0.116573119254879\\
   0.018043066683797   0.072604509112489\\
   0.018043792649143   0.072600089714293\\
   0.018106920143348   0.072213862560532\\
   0.023595896808368   0.038462879577266\\
   0.023596261370248   0.038460070568012\\
   0.023627847325318   0.038215436556051\\
   0.026364235883706   0.016911930224764\\
   0.026376977076731   0.016779627729080\\
   0.027474222591930   0.005296642277619\\
   0.027477358058877   0.005236796551065\\
   0.027742051270691   0.000065788372291\\
   0.027741196337047   0.000045803952192\\
}--cycle;
\coordinate (ptd) at (axis description cs:0,1);
\node[fill=white,align=center,anchor=north west] at (ptd) {\tiny$i=4$};
\coordinate (pt1) at (axis cs:0,0);
\coordinate (ptl2) at (axis description cs:1,0);
\end{axis}

\begin{axis}[%
width=1cm,
height=1cm,
at={(0,0)},
scale only axis,
xmin=-0.2,
xmax=0.2,
xtick={-0.2,0,0.2},
xticklabels={,,},
ymin=-0.2,
ymax=0.2,
ytick={-0.2,0,0.2},
yticklabels={,,},
xshift=6cm,yshift=2.7cm,
minor x tick num={1},
minor y tick num={1},
axis background/.style={fill=white},
]
\addplot[area legend,solid,line width=1.0pt,draw=black,pattern=custom north east lines,forget plot]
table[row sep=crcr] {%
	x	y\\
	-2	-8\\
	2	-8\\
	2	8\\
	-2	8\\
}--cycle;

\addplot[area legend,solid,line width=0.5pt,draw=black,fill=white,forget plot,postaction={pattern=north west lines}]
table[row sep=crcr] {%
	x	y\\
   0.033363073185361   0.000008128621428\\
   0.021018810616548   0.045639089290155\\
   0.013562819449701   0.062136162581653\\
   0.008756084623581   0.072613808369807\\
   0.005652951900109   0.079375862857163\\
   0.003649573198847   0.083741408902742\\
   0.002356193583112   0.086559801407912\\
   0.001521188773532   0.088379353271889\\
   -0.001200857431656   0.094310940957820\\
   -0.001885558224860   0.095802966620624\\
   -0.002924778220519   0.098067522055058\\
   -0.004534476680283   0.101575202180993\\
   -0.007027816801284   0.107008413258920\\
   -0.010889856412426   0.115423766217299\\
   -0.016870732959596   0.128427798694454\\
   -0.026042963239301   0.146251846071511\\
   -0.033363073185361  -0.000008128621428\\
   -0.021018810616548  -0.045639089290155\\
   -0.013562819449701  -0.062136162581653\\
   -0.008756084623581  -0.072613808369807\\
   -0.005652951900109  -0.079375862857163\\
   -0.003649573198847  -0.083741408902742\\
   -0.002356193583112  -0.086559801407912\\
   -0.001521188773532  -0.088379353271889\\
   0.001214638655472  -0.094340971505180\\
   0.001885558224860  -0.095802966620624\\
   0.002924778220519  -0.098067522055058\\
   0.004534476680283  -0.101575202180993\\
   0.007027816801284  -0.107008413258920\\
   0.010889856412426  -0.115423766217299\\
   0.016870732959596  -0.128427798694454\\
   0.026042963239301  -0.146251846071511\\
}--cycle;
\coordinate (pt2) at (axis description cs:0,1);
\end{axis}

\node[square node, name=z1] at (pt1) {};
\draw [-latex] (z1.east)  to node[above,draw=black,fill=white,inner sep=1pt,yshift=-0.02cm,xshift=0.65cm]{\tiny$\times 60$} (pt2);
\draw [-] ($(ptl1)+(0,-0.35cm)$)  to ($(ptl2)+(0,-0.35cm)$);
\draw [-] ($(ptl1)+(-0.6cm,0)$)  to ($(ptl3)+(-0.6cm,0)$);
\end{tikzpicture}%
\caption{\ref{plot:admX} Constraint set $\mathbb{X}_{i}$, \ref{plot:admZ} RPI set $\mathbb{Z}_{i}$.} 
\label{fig.2}
\end{center}
\end{figure}

Suppose first that the following homogeneous local constraints are enforced for all $i\in\mathcal{M}$,
\begin{subequations} 
	\begin{alignat*}{1}
	\mathbb{X}_{i}&=\left\{x^{i}\:\Big\vert\:\begin{bmatrix}
	-2 \\
	-8
	\end{bmatrix}\leq x^{i} \leq \begin{bmatrix}
	2 \\
	8
	\end{bmatrix}\right\}\\
	\mathbb{U}_{i}&=\left\{u^{i}\:\vert\:-4\leq u^{i}\leq 4\right\}.
	\end{alignat*}
\end{subequations}
Table \ref{tb.1} shows a collection of locally stabilizing linear feedbacks, which produce admissible tubes. These local gains correspond to the LQR feedback obtained for $\left(A_{ii},B_{ii}\right)$, $Q_{i}=I_{2}$ and $R_{i}=1$. Figure \ref{fig.2} shows that the RPI set $\mathbb{Z}_{i}$ is contained inside the constraint set $\mathbb{X}_{i}$ for all $i\in\mathcal{M}$ (the inclusion $K_{i}\mathbb{Z}_{i}\subset\mathbb{U}_{i}$ is also met). In view of Theorem \ref{thm.2} it must be then, that $\bm{K}$ is globally stabilizing. This is indeed the case, with $\rho\left(\bm{F}\right)=0.572$.

\begin{table}[hb]
\begin{center}
\caption{Tube admissible local linear feedbacks}
\label{tb.1}
\begin{tabular}{ccccc}
	Feedback & $K^{\top}_{1}$ & $K^{\top}_{2}$ & $K^{\top}_{3}$ & $K^{\top}_{4}$ \\
	\hline
	\rule{0pt}{4ex} Value & $\begin{bmatrix}-1.203\\-0.283\end{bmatrix}$ & $\begin{bmatrix}-0.949\\-0.203\end{bmatrix}$ & $\begin{bmatrix}-1.188\\-0.303\end{bmatrix}$ & $\begin{bmatrix}-1.612\\-0.482\end{bmatrix}$ \\
	\hline
	\rule{0pt}{3ex} $\rho\left(F_{ii}\right)$ & $0.423$ & $0.328$ & $0.422$ & $0.572$ \\
	\hline
\end{tabular}
\end{center}
\end{table}

\subsection{Case 2: Large constraints for a single subsystem}
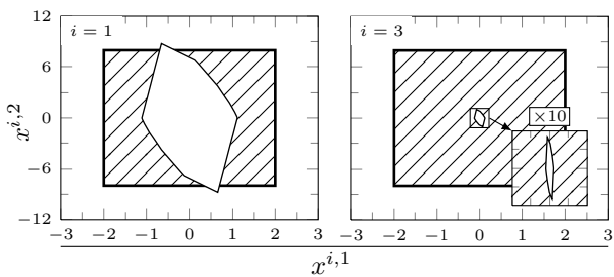
\begin{figure}
\begin{center}
\setlength\figureheight{0.149\textwidth}
\setlength\figurewidth{0.4\textwidth}
\newcommand{\picwidth}{0.47}
\newcommand{\picwidthh}{0.53}

\begin{tikzpicture}[align=center,
square node/.style={rectangle,draw=black,fill=none,thin,minimum width=2.5mm,minimum height=2.5mm},
every pin/.style={fill=white}]
\begin{axis}[%
width=\picwidth\figurewidth,
height=\figureheight,
at={(0\figurewidth,1\figureheight)},
scale only axis,
scale only axis,
xmin=-3,
xmax=3,
xtick={-3,-2,...,3},
ymin=-12,
ymax=12,
ytick={-12,-6,...,12},
minor x tick num={1},
minor y tick num={1},
yticklabel style = {font=\tiny},
xticklabel style = {font=\tiny},
ylabel={$x^{i,2}$},
y label style={at={(-0.3cm,1.3cm)}},
xlabel={$x^{i,1}$},
x label style={at={(3.6cm,-0.3cm)}},
]
\addplot[area legend,solid,line width=1.0pt,draw=black,pattern=custom north east lines,forget plot]
table[row sep=crcr] {%
	x	y\\
	-2	-8\\
	2	-8\\
	2	8\\
	-2	8\\
}--cycle;
\label{plot:nadmX}
\addplot[area legend,solid,line width=0.5pt,draw=black,fill=white,forget plot,postaction={pattern=north west lines}]
table[row sep=crcr] {%
	x	y\\
   1.105344082280190  -0.000488725781458\\
   1.105346399997240  -0.000538275035010\\
   0.656485026963636  -8.764208093068721\\
   0.655990955659092  -8.772802388280260\\
  -0.123703196526727  -6.849615088326270\\
  -0.124510742929005  -6.847515012396120\\
  -0.652921499996090  -3.780928334832070\\
  -0.653463575036453  -3.777753932896710\\
  -0.926638852325907  -1.727666106897140\\
  -0.926917718048760  -1.725563461254690\\
  -1.045755362992650  -0.673147746716531\\
  -1.045876201481940  -0.672073565895905\\
  -1.089837985060830  -0.216814968831732\\
  -1.089882492126660  -0.216352109814685\\
  -1.102969050180590  -0.048991645606313\\
  -1.102982204950510  -0.048822240506545\\
  -1.105344082280190   0.000488725781458\\
  -1.105346399997240   0.000538275035009\\
  -0.656485026963636   8.764208093068721\\
  -0.655990955659092   8.772802388280260\\
   0.123703196526727   6.849615088326270\\
   0.124510742929005   6.847515012396120\\
   0.652921499996090   3.780928334832070\\
   0.653463575036453   3.777753932896710\\
   0.926638852325907   1.727666106897140\\
   0.926917718048760   1.725563461254690\\
   1.045755362992650   0.673147746716531\\
   1.045876201481940   0.672073565895905\\
   1.089837985060830   0.216814968831732\\
   1.089882492126660   0.216352109814684\\
   1.102969050180590   0.048991645606313\\
   1.102982204950510   0.048822240506545\\
}--cycle;
\label{plot:nadmZ}
\coordinate (pta) at (axis description cs:0,1);
\node[fill=white,align=center,anchor=north west] at (pta) {\tiny$i=1$};
\coordinate (ptl1) at (axis description cs:0,0);
\end{axis}

\begin{axis}[%
width=\picwidth\figurewidth,
height=\figureheight,
at={(\picwidthh\figurewidth,1\figureheight)},
scale only axis,
scale only axis,
xmin=-3,
xmax=3,
xtick={-3,-2,...,3},
ymin=-12,
ymax=12,
ytick={-12,-6,...,12},
yticklabels={,,},
minor x tick num={1},
minor y tick num={1},
yticklabel style = {font=\tiny},
xticklabel style = {font=\tiny},
]
\addplot[area legend,solid,line width=1.0pt,draw=black,pattern=custom north east lines,forget plot]
table[row sep=crcr] {%
	x	y\\
	-2	-8\\
	2	-8\\
	2	8\\
	-2	8\\
}--cycle;
\addplot[area legend,solid,line width=0.5pt,draw=black,fill=white,forget plot,postaction={pattern=north west lines}]
table[row sep=crcr] {%
	x	y\\
   0.123804762222301  -0.000987499077658\\
   0.074464190650207  -0.984314010267648\\
   0.074375925841772  -0.984098420808426\\
  -0.013167876151850  -0.768130267347798\\
  -0.013227989250003  -0.767783497229134\\
  -0.072745481186890  -0.423888912684927\\
  -0.072776651109436  -0.423656750452320\\
  -0.103610359754112  -0.193806031634934\\
  -0.103623963673649  -0.193686699778980\\
  -0.117071728252783  -0.075645876907794\\
  -0.117076785716104  -0.075594156109260\\
  -0.122072348446130  -0.024468621712168\\
  -0.122073869110450  -0.024449541166435\\
  -0.123574075655238  -0.005603287836183\\
  -0.123853918001751  -0.000013609089126\\
  -0.123804762222301   0.000987499077658\\
  -0.074464190650207   0.984314010267648\\
  -0.074375925841772   0.984098420808426\\
   0.013167876151850   0.768130267347799\\
   0.013227989250003   0.767783497229134\\
   0.072745481186890   0.423888912684927\\
   0.072776651109436   0.423656750452320\\
   0.103610359754112   0.193806031634934\\
   0.103623963673649   0.193686699778980\\
   0.117071728252782   0.075645876907794\\
   0.117076785716104   0.075594156109260\\
   0.122072348446130   0.024468621712168\\
   0.122073869110450   0.024449541166435\\
   0.123574075655238   0.005603287836183\\
   0.123853918001751   0.000013609089126\\
}--cycle;
\coordinate (ptb) at (axis description cs:0,1);
\node[fill=white,align=center,anchor=north west] at (ptb) {\tiny$i=3$};
\coordinate (ptl2) at (axis description cs:1,0);
\coordinate (pt3) at (axis cs:0,0);
\end{axis}

\begin{axis}[%
width=1cm,
height=1cm,
at={(0,0)},
scale only axis,
xmin=-1.2,
xmax=1.2,
xtick={-1.2,0,1.2},
xticklabels={,,},
ymin=-1.2,
ymax=1.2,
ytick={-1.2,0,1.2},
yticklabels={,,},
xshift=6cm,yshift=2.9cm,
minor x tick num={1},
minor y tick num={1},
axis background/.style={fill=white},
]
\addplot[area legend,solid,line width=1.0pt,draw=black,pattern=custom north east lines,forget plot]
table[row sep=crcr] {%
	x	y\\
	-2	-8\\
	2	-8\\
	2	8\\
	-2	8\\
}--cycle;

\addplot[area legend,solid,line width=0.5pt,draw=black,fill=white,forget plot,postaction={pattern=north west lines}]
table[row sep=crcr] {%
	x	y\\
   0.123804762222301  -0.000987499077658\\
   0.074464190650207  -0.984314010267648\\
   0.074375925841772  -0.984098420808426\\
  -0.013167876151850  -0.768130267347798\\
  -0.013227989250003  -0.767783497229134\\
  -0.072745481186890  -0.423888912684927\\
  -0.072776651109436  -0.423656750452320\\
  -0.103610359754112  -0.193806031634934\\
  -0.103623963673649  -0.193686699778980\\
  -0.117071728252783  -0.075645876907794\\
  -0.117076785716104  -0.075594156109260\\
  -0.122072348446130  -0.024468621712168\\
  -0.122073869110450  -0.024449541166435\\
  -0.123574075655238  -0.005603287836183\\
  -0.123853918001751  -0.000013609089126\\
  -0.123804762222301   0.000987499077658\\
  -0.074464190650207   0.984314010267648\\
  -0.074375925841772   0.984098420808426\\
   0.013167876151850   0.768130267347799\\
   0.013227989250003   0.767783497229134\\
   0.072745481186890   0.423888912684927\\
   0.072776651109436   0.423656750452320\\
   0.103610359754112   0.193806031634934\\
   0.103623963673649   0.193686699778980\\
   0.117071728252782   0.075645876907794\\
   0.117076785716104   0.075594156109260\\
   0.122072348446130   0.024468621712168\\
   0.122073869110450   0.024449541166435\\
   0.123574075655238   0.005603287836183\\
   0.123853918001751   0.000013609089126\\
}--cycle;
\coordinate (pt4) at (axis description cs:0,1);
\end{axis}

\node[square node, name=z2] at (pt3) {};
\draw [-latex] (z2.east)  to node[above,draw=black,fill=white,inner sep=1pt,yshift=-0.02cm,xshift=0.65cm]{\tiny$\times 10$} (pt4);

\draw [-] ($(ptl1)+(0,-0.35cm)$)  to ($(ptl2)+(0,-0.35cm)$);
\end{tikzpicture}
\caption{\ref{plot:admX} Constraint set $\mathbb{X}_{i}$, \ref{plot:admZ} new RPI set $\mathbb{Z}^{new}_{i}$.} 
\label{fig.3}
\end{center}
\end{figure}

Suppose now that, by any reason, truck 2 is allowed thrice the size of the constraints of the other subsystems, i.e. $\mathbb{X}^{\text{new}}_{2}=3\mathbb{X}_{2}$. This effectively means $\mathbb{W}^{\text{new}}_{1}=3\mathbb{W}_{1}$. Although, this new disturbance set is still contained in the state constraint set $\mathbb{X}_{1}$, Fig. \ref{fig.3} shows that the same $K_{1}$ from Table \ref{tb.1} produces a RPI set $\mathbb{Z}^{\text{new}}_{1}$ that is not inside $\mathbb{X}_{1}$ (the RPI set for subsystem 3 also increases in size). This implies that, although the collection of local feedbacks is indeed globally stabilizing, Theorem \ref{thm.2} cannot be used to guarantee it.

\subsection{Case 3: Non-existence}
Heterogeneous constraints are not the only source of conservatism that Theorem \ref{thm.2} suffers from. As made explicit by Remark \ref{rem.1}, a necessary condition for the tubes to be system-wide admissible is that a square PI set exists. Consider the following global system  with $M=2$ (scalar subsystems):
\begin{equation*} 
	A=\begin{bmatrix}
	2.070 & 1.924 \\
	0.316 & 0.203
	\end{bmatrix}, B=\begin{bmatrix}
	0.660 & -1.274 \\
	0.113 & 0.810
	\end{bmatrix}.
\end{equation*}
When the loop is closed with $K_{11}=-2.924$ and $K_{22}=0.977$, Assumption \ref{ass.2} is fulfilled with $\rho\left(F_{11}\right)=0.140$, $\rho\left(F_{22}\right)=0.994$ and $\rho\left(\bm{F}\right)=0.983$. However, it is easy to show that there is no \emph{square} PI set for the closed-loop $\bm{F}$, and therefore, no matter the constraints, the tubes corresponding to the block-diagonal linear feedback will never be simultaneously admissible.

\section{Conclusion} \label{sec.6}
The properties of many distributed MPC techniques usually rely on structured stabilizability assumptions, and standard notions of invariance. In this paper, a fundamental relation between these two concepts has been made explicit. Theorem \ref{thm.2} shows that the constraint admissibility of very simple local robust controllers is a sufficient condition for the global plant to be block-diagonal stabilizable. The main source of conservatism for Theorem \ref{thm.2} is the structure requirement. A necessary condition for the tubes to be system-wide feasible, is that the closed-loop system accepts a \emph{square} PI. Future work will be focused on finding conditions over the coupling terms under which such a particular type of PI set exist.


\end{document}